\documentclass[a4paper,twoside]{article}
\usepackage{amsmath,amssymb,amsfonts}

\usepackage{graphics}
\newtheorem{theorem}{Theorem}[section]

\def\R{{\mathbb R}}

\date{\small\em Version date \today}

\begin{document}
\title{A note on quantum chaology and gamma approximations to eigenvalue spacings
for infinite random matrices}
\author{C.T.J. Dodson \\
{\small School of Mathematics}\\
{\small University of Manchester} \\
{\small Manchester M13 9PL, UK}   }

\maketitle

\begin{abstract}
Quantum counterparts of certain simple classical systems can exhibit chaotic behaviour
through the statistics of their energy levels and the irregular spectra of chaotic systems are modelled by eigenvalues of infinite random matrices.
We use known bounds on the distribution
function for eigenvalue spacings for the Gaussian orthogonal ensemble (GOE) of infinite random real symmetric matrices and show that gamma distributions, which have an important uniqueness property, can yield an approximation to the GOE distribution. That has
the advantage that then both chaotic and non chaotic cases fit in the
information geometric framework of the manifold of gamma distributions, which has been
the subject of recent work on neighbourhoods of randomness for general stochastic systems.
Additionally, gamma
distributions give approximations, to eigenvalue spacings for the Gaussian unitary ensemble (GUE) of infinite random hermitian matrices and for the Gaussian symplectic ensemble (GSE) of infinite random hermitian matrices with real quaternionic elements, except near the origin. Gamma distributions do not
precisely model the various analytic systems discussed here, but some features may be useful in studies of qualitative generic properties in applications to data from real systems which manifestly seem to exhibit behaviour reminiscent of near-random processes. 
\end{abstract}

\section{Introduction}
Berry introduced the term quantum chaology in his 1987 Bakerian Lecture~\cite{berry87}
as the study of semiclassical but non-classical behaviour of systems whose classical
motion exhibits chaos. He illustrated it with the statistics of energy levels, following
his earlier work with Tabor~\cite{berrytabor77} and related developments from the study
of a range of systems. In the regular spectrum of a bound system with $n\geq 2$ degrees
of freedom and $n$ constants of motion, the energy levels are labelled by $n$
quantum numbers, but the quantum numbers
of nearby energy levels may be very different. In the case of an irregular spectrum, such
as for an ergodic system where only energy is conserved, we cannot use quantum number labelling.
This prompted the use of energy level spacing distributions to allow comparisons
among different spectra~\cite{berrytabor77}. It was known, eg from the work of Porter~\cite{porter},
that the spacings between energy levels of complex nuclei and atoms with $n$ large are modelled
by the spacings of eigenvalues of random matrices and that the Wigner distribution~\cite{wigner67}  gives a very good fit. It turns out that the spacing distributions
for generic regular systems are negative exponential, that is random; but for irregular systems
the distributions are skew and unimodal, at the scale of the mean spacing.
Mehta~\cite{mehta} provides a detailed discussion of
the numerical experiments and functional approximations to the energy level spacing statistics,
Alt et al~\cite{alt} compare eigenvalues from numerical analysis and from microwave resonator experiments, also eg. Bohigas et al~\cite{bohigas84} and Soshnikov~\cite{soshnikov} confirm certain universality properties. Also, Miller~\cite{miller06} provides much detail on a range of related number theoretic properties, including random matrix theory links with L-functions. Forrester's online book~\cite{forrester} gives a wealth of analytic detail on the mathematics and physics of eigenvalues of infinite random matrices for the three ensembles of particular interest: Gaussian orthogonal (GOE), unitary (GUE) and symplectic (GSE), being the real, complex and quaternionic cases, respectively. The review by Deift~\cite{deift} illustrates how random matrix theory has significant links to a wide range of mathematical problems in the theory of functions as well as to mathematical physics. From~\cite{forrester}, we have definitions for the three cases of interest:
\begin{description}
\item[GOE:] A random real symmetric $n\times n$ matrix belongs to the Gaussian orthogonal ensemble (GOE) if the diagonal and upper triangular elements are independent random variables with Gaussian distributions of  zero mean and standard deviation $1$ for the diagonal and $\frac{1}{\sqrt{2}}$ for the upper triangular elements.
\item[GUE:] A random hermitian $n\times n$ matrix belongs to the Gaussian unitary ensemble (GUE) if the diagonal elements $m_{jj}$ (which must be real) and the upper triangular elements $m_{jk}=u_{jk}+ i v_{jk}$ are independent random variables with Gaussian distributions of  zero mean and standard deviation $\frac{1}{\sqrt{2}}$ for the $m_{jj}$ and $\frac{1}{2}$ for each of the $u_{jk}$ and $v_{jk}.$
\item[GSE:] A random hermitian $n\times n$ matrix with real quaternionic elements belongs to the Gaussian symplectic ensemble (GSE) if the diagonal elements $z_{jj}$ (which must be real) are independent with Gaussian distribution of zero mean and standard deviation $\frac{1}{2}$ and the upper triangular elements $z_{jk}=u_{jk}+ i v_{jk}$ and  $w_{jk}=u'_{jk}+ i v'_{jk}$ are independent random variables with Gaussian distributions of  zero mean and standard deviation $\frac{1}{2\sqrt{2}}$ for each of the $u_{jk}, \ u'_{jk}, \ v'_{jk}$ and $v_{jk}.$
\end{description}
Then the matrices in these ensembles are respectively invariant under the appropriate orthogonal, unitary and symmetric transformation groups, and moreover in each case the joint density function of all independent elements is controlled by the trace of the matrices and is of form~\cite{forrester}
\begin{equation}\label{pdftrace}
    p(X) = A_n \, e^{-\frac{1}{2}TrX^2}
\end{equation}
where $A_n$ is a normalizing factor. Barndorff-Nielsen et al~\cite{BNGJ} give some background mathematical statistics on the more general problem of quantum information and quantum statistical inference, including reference to random matrices.

Here we show that gamma distributions provide approximations to eigenvalue spacing distributions for the GOE distribution comparable to the Wigner distribution at the scale of the mean and for the GUE and GSE distributions, except near the origin. That may be useful because the gamma distribution has a well-understood and tractable information geometry~\cite{gamran,vpf06} as well as the following important uniqueness property:
\begin{theorem}[Hwang and Hu~\cite{hwang}]\label{hwangthm}
For independent positive random variables with a common probability density function $f,$
having independence of the sample mean and the sample coefficient of variation is
equivalent to $f$ being the gamma distribution.
\end{theorem}
 It is noteworthy also that the non-chaotic case has an exponential distribution of spacings between energy levels and that the sum
of $n$ independent identical exponential random variables follows
a gamma distribution and moreover the sum
of $n$ independent identical gamma random variables follows
a gamma distribution; furthermore, the product of gamma distributions is well-approximated by a gamma distribution.

\begin{figure}
\begin{center}
\begin{picture}(300,200)(0,0)
\put(0,0){\resizebox{10cm}{!}{\includegraphics{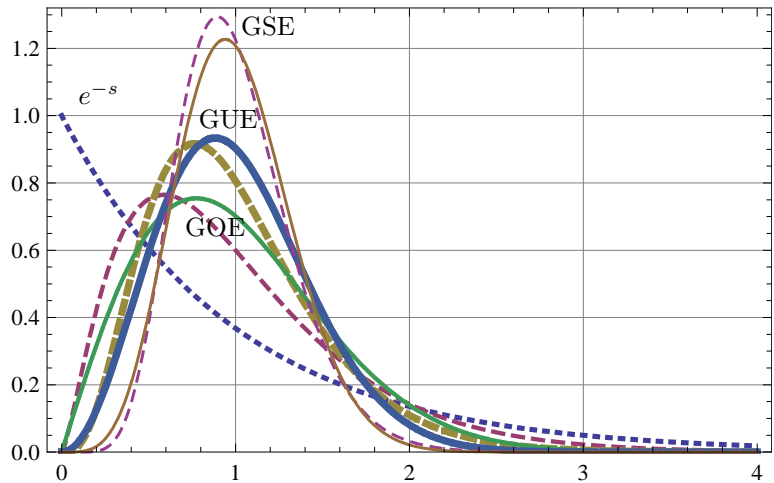}}}
\put(86,170){GSE}
\put(70,135){GUE}
\put(65,94){GOE}
\put(25,144){$e^{-s}$}
\end{picture}
\end{center}
\caption{{\em Probability density functions for the unit mean gamma distributions (dashed) and generalized gamma distribution (solid) fits to the true variances for left to right the GOE , GUE and GSE cases. The two types coincide in the exponential case, $e^{-s},$ shown dotted.}}
\label{GGamGamFits}
\end{figure}

 From a different standpoint, Berry and Robnik~\cite{berryrobnik84} gave a statistical model using a  mixture of energy level spacing sequences from exponential and Wigner distributions. Monte Carlo methods were used by Ca\"{e}r et al.~\cite{caer} to investigate such a mixture. Ca\"{e}r et al. established also the best fit of GOE, GUE and GSE unit mean distributions, for spacing $s>0,$ using the generalized gamma density\index{generalized gamma distribution} which we can put in the form
 \begin{eqnarray}\label{ggamma}
    g(s; \beta, \omega) &=& a(\beta, \omega) \, s^\beta \, e^{-b(\beta, \omega) s^\omega } \ \ {\rm for} \ \beta, \omega > 0\\
    {\rm where} \ \ a(\beta, \omega) &=& \frac{ \omega\left[ \Gamma((2+\beta)/\omega)\right]^
    {\beta+1}}{\left[\Gamma((1+\beta)/\omega)\right]^{\beta+2} }\ \ {\rm and} \ \
    b(\beta, \omega)= \left[\frac{\Gamma((2+\beta)/\omega) }
    { \Gamma((1+\beta)/\omega)}\right]^\omega \ . \nonumber
 \nonumber
 \end{eqnarray}
Then the best fits of (\ref{ggamma}) had the parameter values~\cite{caer}
\begin{center}
\begin{tabular}{|c||c|c|c|}
  \hline
  Ensemble & $\beta$ & $\omega$ & Variance\\ \hline\hline
  GOE & 1 & 1.886 &0.2856\\
  GUE & 2 & 1.973 &0.1868\\
  GSE & 4 & 2.007 &0.1100\\
  \hline
\end{tabular}
\end{center}\label{gengamapprox}
and were accurate to within $\sim 0.1\%$ of the true distributions from Forrester~\cite{forrester}. Observe that the exponential distribution is recovered by the choice $g(s; 0, 1)=e^{-s}.$ These  distributions are shown in Figure~\ref{GGamGamFits} along with corresponding fits of the gamma distribution.

\section{Eigenvalues of Random Matrices}
The two classes of spectra are illustrated in two dimensions by bouncing geodesics in
 plane billiard tables: eg in the de-symmetrized `stadium of Bunimovich' with ergodic chaotic behaviour
 and irregular spectrum on the one hand, and on the other hand
in the symmetric annular region between concentric circles
 with non-chaotic behaviour, regular spectrum and random energy spacings~\cite{berrytabor77,bohigas84,mehta,berry87}.

 \begin{figure}
\begin{center}
\begin{picture}(300,200)(0,0)
\put(0,0){\resizebox{10cm}{!}{\includegraphics{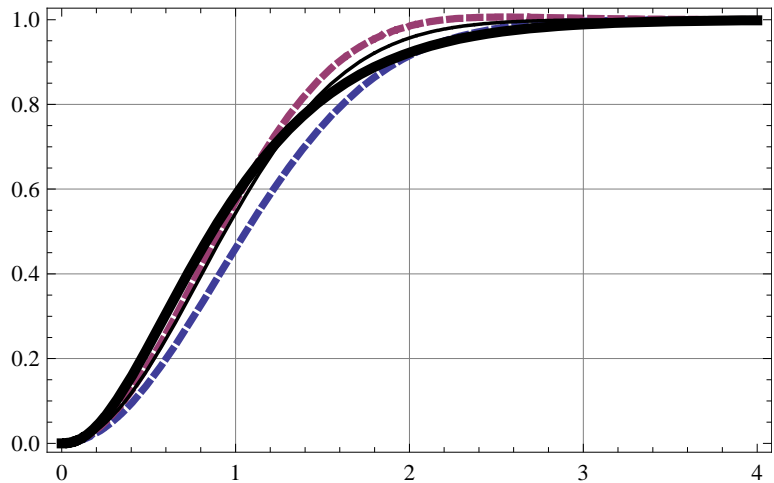}}}
\end{picture}
\end{center}
\caption{{\em  The bounds on the normalized cumulative distribution function of eigenvalue spacings for the GOE of random matrices
\protect{(\ref{bounds})} (dashed), the Wigner surmise \protect{(\ref{W})} (thin solid) and the unit mean gamma
distribution fit to the true GOE distribution from Mehta~\cite{mehta} Appendix A.15 (thick solid).}}
\label{cdfs}
\end{figure}
It turns out that the mean spacing between eigenvalues of infinite symmetric real random
matrices---the so called Gaussian Orthogonal Ensemble (GOE)---is bounded and therefore it is convenient to normalize the distribution to have unit mean; also, the same is true for the GUE and GSE cases.  Barnett~\cite{barnett} provides a numerical tabulation of the first 1,276,900 eigenvalues. In fact, Wigner~\cite{wigner55,wigner58,wigner67}
had already surmised that the cumulative probability distribution function at the scale of the mean spacing should be of the form:
\begin{equation}\label{W}
    W(s) = 1-e^{-\frac{\pi  s^2}{4}}
\end{equation}
which has unit mean and variance $\frac{4-\pi}{\pi}\approx 0.273$ with probability density function
\begin{equation}\label{w}
    w(s) = \frac{\pi}{2} s \, e^{-\frac{\pi  s^2}{4}}.
\end{equation}
Remarkably, Wigner's surmise gave an extremely good fit with numerical computation of the true GOE distribution, cf. Mehta~\cite{mehta} Appendix A.15, and with a variety of observed data from atomic and nuclear systems~\cite{wigner67,berrytabor77,berry87,mehta}.

From Mehta~\cite{mehta} p 171, we have bounds on the cumulative probability distribution
function $P$ for the spacings between eigenvalues of infinite symmetric real random matrices:
\begin{equation}\label{bounds}
    L(s)=1-e^{-\frac{1}{16} \pi ^2 s^2}\leq \ P(s) \ \leq U(s)=1-e^{-\frac{1}{16} \pi ^2 s^2} \left(1-\frac{\pi ^2 s^2}{48}\right).
\end{equation}
Here the lower bound $L$ has mean $\frac{2}{\sqrt{\pi}}\approx 1.13$ and
variance $\frac{4(4-\pi)}{\pi^2}\approx 0.348,$  and the upper bound $U$
has mean $\frac{5}{3\sqrt{5}}\approx 0.940$ and
variance $\frac{96-25\pi}{9\pi^2}\approx 0.197.$

\begin{figure}
\begin{center}
\begin{picture}(300,200)(0,0)
\put(0,0){\resizebox{10cm}{!}{\includegraphics{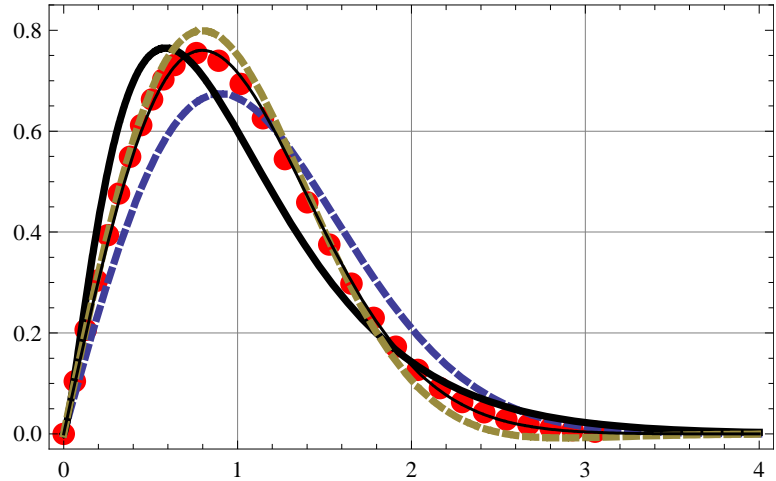}}}
\end{picture}
\end{center}
\caption{{\em Probabilty density function for the unit mean gamma distribution  fit (thick solid) to the true GOE distribution from Mehta~\cite{mehta} Appendix A.15 (points),  the Wigner surmised density  \protect{(\ref{w})} (thin solid) and the probability densities for the bounds \protect{(\ref{pbounds})} (dashed) for the distribution of normalized spacings between eigenvalues for infinite symmetric real random matrices.}}
\label{pdfs}
\end{figure}

The family of probability density functions for gamma distributions with dispersion parameter $\kappa >0$ and mean $\kappa/\nu >0$ for positive random variable $s$ is given by
\begin{equation}\label{gammapdfs}
    f(s;\nu,\kappa) = \nu^\kappa  \frac{s^{\kappa-1}}{\Gamma(\kappa)} e^{-s\nu}, \ \ {\rm for} \ \nu,\kappa>0
\end{equation}
with variance $\frac{\kappa}{\nu^2}.$
Then the subset having unit mean is given by
\begin{equation}\label{unimeangamma}
    f(s;\kappa,\kappa) = \kappa^\kappa \frac{s^{\kappa-1}}{\Gamma(\kappa)} e^{-s\kappa}, \ \ {\rm for} \ \kappa>0
\end{equation}
with variance $\frac{1}{\kappa}.$ These parameters $\nu,\kappa$ are called natural parameters because they admit presentation of the family (\ref{gammapdfs}) as an exponential family~\cite{AN} and thereby provide an associated natural affine immersion in $\R^3$~\cite{affimm}
\begin{equation}\label{afim}
    h: \R^+\times\R^+ \rightarrow \R^3 :
  \left( \! \!
    \begin{array}{c}
       \nu \\ \kappa
    \end{array} \! \! \right)
    \mapsto
    \left( \! \!
    \begin{array}{c}
       \nu \\ \kappa \\ \log\Gamma(\kappa) - \kappa\log\nu
    \end{array} \! \! \right).
\end{equation}
This affine immersion was used~\cite{gamran} to present tubular neighbourhoods of the 1-dimensional
subspace consisting of exponential distributions $(\kappa=1),$ so giving neighbourhoods of random processes.
The maximum entropy case has $\kappa=1$ and corresponds to
an underlying Poisson random event process and so models spacings in the spectra for non-chaotic systems; for $\kappa>1$ the distributions are skew unimodular.
The unit mean gamma distribution fit to the true GOE distribution from Mehta~\cite{mehta} has
variance $\approx 0.379$ and hence $\kappa\approx 2.42.$

In fact, $\kappa$ is a geodesic coordinate in the Riemannian 2-manifold of gamma distributions
with Fisher information metric~\cite{gamran}; arc length along this coordinate from
$\kappa=a$ to $\kappa=b$ is given by
\begin{equation}\label{garclength}
    \left|\int_a^b \sqrt{\frac{d^2 \log(\Gamma(\kappa))}{d\kappa^2} -\frac{1}{\kappa}} \, d\kappa\right| \, .
\end{equation}

 Plotted in Figure~\ref{cdfs} are the cumulative distributions for the bounds (\ref{bounds}) (dashed), the gamma distribution (thick solid) fit to the true GOE distribution with unit mean, and the Wigner surmised distribution (\ref{W}) (thin solid).
The corresponding probability density functions are in Figure~\ref{pdfs}:  unit mean gamma distribution  fit (thick solid) to the true GOE distribution from Mehta~\cite{mehta} Appendix A.15 (points),  the Wigner surmised density  \protect{(\ref{w})} (thin solid) and the probability densities for the bounds \protect{(\ref{pbounds})} (dashed), respectively,
\begin{equation}\label{pbounds}
  l(s)= \frac{\pi  s}{2} e^{-\frac{\pi  s^2}{4}}  , \ \ \ u(s)=\frac{\pi ^2 s(64-\pi ^2 s^2)}{384} e^{-\frac{1}{16}\pi ^2 s^2}.
\end{equation}

\section{Deviations}
Berry has pointed out~\cite{berry08} that the behaviour near the origin is an important feature of the ensemble statistics of these matrices and in particular the GOE distribution is linear near the origin, as is the Wigner distribution. Moreover,
for the unitary ensemble (GUE) of complex hermitian matrices,
near the origin, the behaviour is $\sim s^2$
and for the symplectic ensemble (GSM, representing half-integer spin particles with
time-reversal symmetric interactions) it is $\sim s^4.$

\begin{figure}
\begin{center}
\begin{picture}(300,250)(0,0)
\put(-0,0){\resizebox{8cm}{!}{\includegraphics{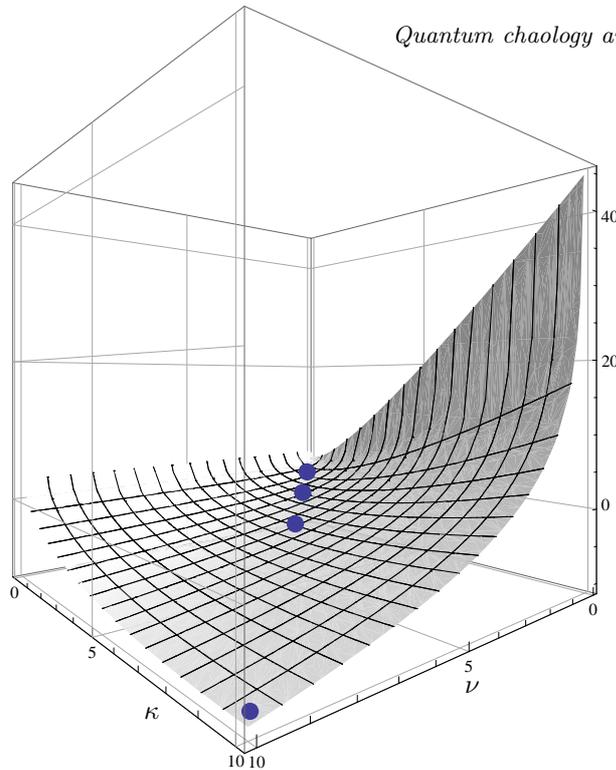}}}
\put(170,30){$\nu$}
 \put(50,20){$\kappa$}
\end{picture}
\end{center}
\caption{{\em The unit mean gamma distributions corresponding to the random (non-chaotic) case, $\kappa=\nu = 1$ and those with exponent $\kappa=\nu =  2.420, \ 4.247, \ 9.606$ for the best fits to the true variances of the spacing distributions for the GOE, GUE and GSE cases, as points on the affine immersion in $\R^3$ of the $2$-manifold of gamma distributions.}} \label{Gamma1249G}
\end{figure}

From (\ref{unimeangamma}) we see that at unit mean the gamma density behaves like $s^{\kappa-1}$ near the origin, so linear behaviour would require $\kappa=2$ which gives a variance of $\frac{1}{\kappa}=\frac{1}{2}$ whereas the GOE fitted gamma distribution has $\kappa\approx 2.42$ and hence variance $\approx 0.379.$ This may be compared with variances for the lower bound $l,$   $\frac{4(4-\pi)}{\pi^2}\approx 0.348,$  the upper bound $u,$ $\frac{96-25\pi}{9\pi^2}\approx 0.197,$ and the Wigner distribution $w,$ $\frac{4-\pi}{\pi}\approx 0.273.$ The gamma distributions fitted to the lower and upper bounding distributions have, respectively, $\kappa_L=\frac{\pi }{4-\pi }\approx 3.660$ and
$\kappa_U=\frac{5\pi^2 }{96-25\pi }\approx 2.826.$  Figure~\ref{GGamGamFits} shows the probability density functions for the unit mean gamma distributions (dashed) and generalized gamma distribution (solid) fits to the true variances for left to right the GOE , GUE and GSE cases; the two types coincide in the exponential case, $e^{-s},$ shown dotted. The major differences are in the behaviour near the origin. Figure~\ref{Gamma1249G} shows unit mean gamma distributions with $\kappa=\nu =  2.420, \ 4.247, \ 9.606$ for the best fits to the true variances of the spacing distributions for the GOE, GUE and GSE cases, as points on the affine immersion in $\R^3$ of the $2$-manifold of gamma distributions, cf.~\cite{affimm}. The information metric provides information distances on the gamma manifold and so could be used for comparison of real data on eigenvalue spacings if fitted to gamma distributions; that may allow identification of qualitative properties and represent trajectories during structural changes of systems.

\begin{figure}
\begin{center}
\begin{picture}(300,200)(0,0)
\put(0,0){\resizebox{10cm}{!}{\includegraphics{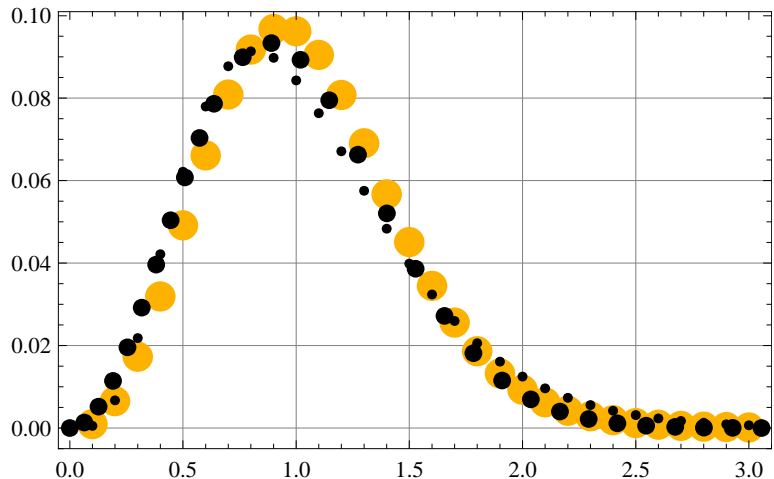}}}
\end{picture}
\end{center}
\caption{{\em Probability plot with unit mean for the spacings between the first 2,001,052 zeros of the Riemann zeta function from the tabulation of Odlyzko~\cite{odlyzko} (large points),  that
for the true GUE distribution from the tabulation of Mehta~\cite{mehta} Appendix A.15 (medium points) and the gamma fit to the true GUE (small points).}}
\label{zz2mguegam}
\end{figure}
The author is indebted to Rudnick~\cite{rudnick08} for pointing out that the GUE eigenvalue spacing
distribution is rather closely followed by the distribution of zeros for the Riemann zeta function; actually, Hilbert had conjectured this, as mentioned along with a variety of other probabilistic aspects of number theory by Schroeder~\cite{schroeder}.  This can be seen in Figure~\ref{zz2mguegam} which shows with unit mean the probability distribution for spacings
among the first 2,001,052 zeros from the tabulation of Odlyzko~\cite{odlyzko} (large points), that
for the true GUE distribution from the tabulation of Mehta~\cite{mehta} Appendix A.15 (medium points) and the gamma fit to the true GUE (small points), which has $\kappa\approx 4.247.$ The grand mean spacing between zeros from the data was $\approx 0.566,$ the coefficient of variation $\approx 0.422$ and variance $\approx 0.178.$

\begin{table}
\begin{center}
$\begin{array}{|c||l|l|c|c|}
\hline
Block&Mean & Variance & CV & \kappa \\
\hline
 1 & 1.232360 & 0.276512 & 0.426697 & 5.49239 \\
 2 & 1.072330 & 0.189859 & 0.406338 & 6.05654 \\
 3 & 1.025210 & 0.174313 & 0.407240 & 6.02974 \\
 4 & 0.996739 & 0.165026 & 0.407563 & 6.02019 \\
 5 & 0.976537 & 0.158777 & 0.408042 & 6.00607 \\
 6 & 0.960995 & 0.154008 & 0.408367 & 5.99651 \\
 7 & 0.948424 & 0.150136 & 0.408544 & 5.99131 \\
 8 & 0.937914 & 0.147043 & 0.408845 & 5.98250 \\
 9 & 0.928896 & 0.144285 & 0.408926 & 5.98014 \\
 10 & 0.921034 & 0.142097 & 0.409276 & 5.96991\\
 \hline
\end{array}$\end{center}\label{cvzt}
\caption{{\em Effect of location: Statistical data for spacings in the first ten consecutive blocks of 200,000 zeros of the Riemann zeta function normalized with unit grand mean from the tabulation of Odlyzko~\cite{odlyzko}.}}
\end{table}

 Table 1 shows the effect of location on the statistical data for spacings in the first ten consecutive blocks of 200,000 zeros of the Riemann zeta function normalized with unit grand mean; Table 2 shows the effect of sample size.  For gamma distributions we expect the coefficient of variation to be independent of sample size and location, by  Theorem~\ref{hwangthm}.

{\bf Remark}
The gamma distribution provides approximations to the true distributions for the spacings between  eigenvalues of infinite random matrices for the GOE, GUE and the GSE cases.
However, it is clear that gamma distributions do not
precisely model the analytic systems discussed here, and do not give correct asymptotic behaviour at the origin, as is evident from the results of Ca\"{e}r et al.~\cite{caer} who obtained excellent approximations for GOE, GUE and GSE distributions using the generalized gamma distribution (\ref{ggamma})  . The differences may be seen in Figure~\ref{GGamGamFits} which shows the unit mean distributions for gamma (dashed) and generalized gamma~\cite{caer} (solid) fits to the true variances for the Poisson, GOE, GUE and GSE ensembles.

The generalized gamma distributions do not have a tractable information geometry and so some features of the gamma distribution approximations may be useful in studies
of qualitative generic properties in applications to data from real systems.
  It would be interesting to investigate the extent to which data from real atomic and nuclear systems has generally the qualitative property that the sample coefficient of variation is independent of the mean. That, by Theorem~\ref{hwangthm}, is an information-theoretic distinguishing property of the gamma distribution.

\begin{table}
\begin{center}
$\begin{array}{|c||l|l|c|c|}
\hline
m &Mean & Variance & CV & \kappa \\
\hline
  1 & 1.23236 & 0.276511 & 0.426696 & 5.49242 \\
 2 & 1.15234 & 0.239586 & 0.424765 & 5.54246 \\
 3 & 1.10997 & 0.221420 & 0.423934 & 5.56421 \\
 4 & 1.08166 & 0.209725 & 0.423384 & 5.57869 \\
 5 & 1.06064 & 0.201303 & 0.423018 & 5.58833 \\
 6 & 1.04403 & 0.194799 & 0.422748 & 5.59548 \\
 7 & 1.03037 & 0.189538 & 0.422527 & 5.60133 \\
 8 & 1.01881 & 0.185161 & 0.422357 & 5.60584 \\
 9 & 1.00882 & 0.181418 & 0.422207 & 5.60983 \\
 10 & 1.00004 & 0.178180 & 0.422094 & 5.61282\\
 \hline
\end{array}$\end{center}\label{cvsst}
\caption{{\em Effect of sample size: Statistical data for spacings in ten blocks of increasing size  $200,000m, \ m=1,2,\ldots,10,$ for the first 2,000,000 zeros of the Riemann zeta function, normalized with unit grand mean,  from the tabulation of Odlyzko~\cite{odlyzko}.}}
\end{table}
  It would be interesting also to know if there is a number-theoretic property that corresponds to the apparently similar qualitative behaviour of the spacings of zeros of the Riemann zeta function, Tables 12.1, 12.2. Since the non-chaotic case has an exponential distribution of spacings between energy levels and the sum
of $n$ independent identical exponential random variables follows
a gamma distribution and moreover the sum
of $n$ independent identical gamma random variables follows
a gamma distribution, a further analytic development would be to calculate the eigenvalue distributions for gamma or loggamma-distributed matrix ensembles. Information geometrically, the Riemannian manifolds of gamma and loggamma families are isometric,  but the loggamma random variables have bounded domain and their distributions contain the uniform distribution, which may be important in modelling some real physical processes.


\begin{thebibliography}{99}

\bibitem{alt} H. Alt, C. Dembrowski, H.D. Graf, R. Hofferbert, H. Rehfield, A. Richter and C. Schmit. Experimental versus numerical eigenvalues of a Bunimovich stadium billiard: A comparison.
    {\em Phys. Rev. E} 60, 3 (1999) 2851-2857.


\bibitem{AN} S-I. Amari  and H. Nagaoka. {\bf Methods of Information Geometry},
 American Mathematical Society, Oxford University Press, 2000.

\bibitem{gamran} Khadiga Arwini and C.T.J. Dodson.
Information geometric neighbourhoods of randomness and geometry of
the McKay bivariate gamma 3-manifold. {\em Sankhya: Indian Journal
of Statistics} 66, 2 (2004) 211-231.


\bibitem{BNGJ} O.E. Barndorff-Nielsen, R.D. Gill and P.E. Jupp. On Quantum Statistical Inference. Preprint (2001).


\bibitem{barnett} A.H. Barnett. \verb+http://math.dartmouth.edu/~ahb/pubs.html+


\bibitem{berry08} M.V. Berry. Private communication. 2008.

\bibitem{berry87} M.V. Berry. Quantum Chaology. {\em Proc. Roy. Soc. London  A} 413, (1987) 183-198.

\bibitem{berrytabor77} M.V. Berry and M. Tabor. Level clustering in the regular spectrum. {\em Proc. Roy. Soc. London  A} 356, (1977) 373-394.

\bibitem{berryrobnik84} M.V. Berry and M. Robnik. Semiclassical level spacings when regular
and chaotic orbits coexist. {\em J. Phys. A Math. General} 17, (1984) 2413-2421.

\bibitem{bohigas84} O. Bohigas, M.J. Giannoni and C. Schmit. Characterization of Chaotic Quantum Spectra
and Universality of Level Fluctuation Laws. {\em Phys. Rev. Lett.} 52, 1 (1984) 1-4.


\bibitem{caer} G. Le Ca\"{e}r, C. Male and R. Delannay. Nearest-neighbour spacing distributions of the $\beta$-Hermite ensemble of random matrices. {\em Physica A} (2007) 190-208. Cf. also their Erratum: {\em Physica A} 387 (2008) 1713.

\bibitem{deift} P. Deift. Some open problems in random matrix theory and the theory of integrable systems. Preprint,
arXiv:arXiv:0712.0849v1 6 December 2007.

\bibitem{vpf06} C.T.J. Dodson. Quantifying galactic clustering and departures from randomness of the inter-galactic void probablity function using information geometry. http://arxiv.org/abs/astro-ph/0608511 (2006).

\bibitem{affimm} C.T.J. Dodson and Hiroshi Matsuzoe.
An affine embedding of the gamma manifold. {\em InterStat},
January 2002, 2 (2002) 1-6.


\bibitem{forrester} P. J. Forrester, Log-Gases and Random Matrices, Chapter 1 Gaussian matrix ensembles. Book manuscript, \verb+http://www.ms.unimelb.edu.au/~matpjf/matpjf.html+, 2007.

\bibitem{hwang} T-Y. Hwang and C-Y. Hu. On a characterization of
the gamma distribution: The independence of the sample mean and
the sample coefficient of variation. {\em Annals Inst. Statist.
Math.} 51, 4 (1999) 749-753.

\bibitem{kullback} S. Kullback. {\bf Information and Statistics},
J. Wiley, New York, 1959.

\bibitem{mehta} Madan Lal Mehta. {\bf Random Matrices} $3^{rd}$ Edition, Academic Press,  London 2004.


 \bibitem{miller06} Steven J. Miller and Ramin Takloo-Bighash. {\bf An Invitation to Modern Number Theory} Princeton University Press, Princeton 2006. Cf. also the seminar notes:
 \begin{itemize}\item Steven J. Miller: Random Matrix Theory, Random
Graphs, and L-Functions: How the Manhatten Project helped
us understand primes. Ohio State University Colloquium 2003.\\
\verb+http://www.math.brown.edu/~sjmiller/math/talks/colloquium7.pdf+ .
 \item Steven J. Miller. Random Matrix Theory Models for zeros
of L-functions near the central point (and applications to elliptic curves). Brown University Algebra Seminar 2004.\\
\verb+http://www.math.brown.edu/~sjmiller/math/talks/RMTandNTportrait.pdf+ .
\end{itemize}


\bibitem{odlyzko} A. Odlyzko. Tables of zeros of the Riemann zeta function. \\
\verb+http://www.dtc.umn.edu:80/~odlyzko/zeta_tables/index.html+ .

\bibitem{porter} C.F. Porter. {\bf Statistical Theory of Spectra: Fluctuations} Edition, Academic Press, London 1965.

\bibitem{rudnick08} Z. Rudnick. Private communication. 2008. Cf. also
Z. Rudnick. What is Quantum Chaos? {\em Notices A.M.S.}  55, 1 (2008) 33-35.

\bibitem{schroeder} M.R. Schroeder. {\bf Number Theory in Science and Communication. With Applications in Cryptography, Physics, Digital Information, Computing, and Self-Similarity}. Springer Series in Information Science, 3$^{rd}$ edition, Springer, Berlin 1999.



\bibitem{soshnikov} A. Soshnikov. Universality at the edge of the spectrum in Wigner random
matrices. {\em Commun. Math. Phys.} 207 (1999) 697-733.

\bibitem{wigner55} E.P. Wigner. Characteristic vectors of bordered matrices with infinite
dimensions. {\em Annals of Mathematics} 62, 3 (1955) 548-564.

\bibitem{wigner58} E.P. Wigner. On the distribution of the roots of certain symmetric
matrices. {\em Annals of Mathematics} 67, 2 (1958) 325-327.

\bibitem{wigner67} E.P. Wigner. Random matrices in physics. {\em SIAM Review} 9, 1 (1967) 1-23.

\end{thebibliography}
\end{document}